# Orientation-tunable correlated Chern insulating states in chiral twisted double bilayer graphene proximitized by $WSe_2$

Jiao Xie[1,2], Yongqin Xie[1], Fanqiang Chen[1], Jiliang Yang[3*], Sicheng Chen[1], Moyu Chen[1], Kenji Watanabe[4], Takashi Taniguchi[5], Shi-Jun Liang[1], Kemi Xu[3,6], Bin Cheng[1,7*], and Feng Miao[1,8*]

[1]Institute of Brain-Inspired Intelligence, National Laboratory of Solid State Microstructures, School of Physics, Collaborative Innovation Center of Advanced Microstructures, Nanjing University, Nanjing 210093, China.

[2]Hefei National Laboratory, Hefei 230088, China.

[3]Yangtze Delta Region Academy of Beijing Institute of Technology (Jiaxing), Jiaxing, 314019, China.

[4]Research Center for Functional Materials, National Institute for Materials Science, Tsukuba, Japan.

[5]International Center for Materials Nanoarchitectonics, National Institute for Materials Science, Tsukuba, Japan.

[6]State Key Laboratory of Environment Characteristics and Effects for Near-space, MIIT Key Laboratory of Complex-field Intelligent Sensing, School of Optics and Photonics, Center for Photonic Quantum Precision Measurement, Beijing Institute of Technology, Beijing, 100081, China.

[7]Institute of Interdisciplinary Physical Sciences, School of Science, Nanjing University of Science and Technology, Nanjing 210094, China.

[8]Jiangsu Physical Science Research Center, Nanjing 210093, China.

*Corresponding author. Email: jlyang@bit.edu.cn; bincheng@njust.edu.cn; miao@nju.edu.cn

**Abstract:** Moiré flat bands in graphene systems proximitized by transition-metal dichalcogenides (TMDCs) provide a setting where spin-orbit coupling (SOC) can reshape band topology. The crystallographic alignment angle $\Phi$ between TMDC and graphene layers is predicted to tune the balance of Ising and Rashba SOC, but a combined theoretical and experimental understanding of how $\Phi$ governs the topological character of correlated states has not been systematically established. Here we show that in chiral-stacked twisted double bilayer graphene in proximity to $WSe_2$, $\Phi$ determines the topological character of correlated Chern insulators. Continuum model calculations reveal that Ising spin-orbit coupling dominates at $\Phi \approx 0°$, giving rise to flat bands with finite valley Chern numbers, whereas Rashba coupling dominates at larger $\Phi$, resulting in topologically trivial bands. Transport measurements at quarter filling confirm this picture: $\Phi \approx 0°$ devices exhibit $C = +1$ Chern insulators, consistent with spontaneous isospin polarization, whereas $\Phi \approx 15°$ devices show $C = 0$ despite exhibiting similar correlated insulating behavior. The sharp contrast establishes crystallographic alignment as an additional tuning knob, complementary to twist angle, displacement field, and carrier density, for engineering correlated topological states in van der Waals heterostructures.

## Introduction

Moiré superlattices created by twisting or lattice mismatch in two-dimensional van der Waals heterostructures provide a highly tunable setting to study how band topology intertwines with strong electron correlations[1-4]. In moiré graphene systems, the flat bands can reach bandwidths comparable to or smaller than the Coulomb energy scale[5,6], enabling correlated insulators[1,7-17], superconductivity[2,7,8,18-20], and spontaneous symmetry breaking[9,10,17,21-25]. A key theoretical insight is that these flat bands often carry nontrivial valley Chern numbers, which arise from momentum-space Berry curvature concentrated near each valley[26-29]. This Berry curvature arises from broken inversion symmetry, and spontaneous flavor polarization yields correlated Chern insulators by breaking time-reversal symmetry, as observed in magic-angle twisted bilayer graphene (MATBG) aligned with hexagonal boron nitride (hBN)[14,30], twisted double bilayer graphene (TDBG)[31], and rhombohedral multilayer graphene/hBN moiré superlattices[4,32].

A complementary and highly flexible route to engineer Berry curvature and valley Chern numbers is proximity-induced spin–orbit coupling (SOC) from transition-metal dichalcogenides (TMDCs)[33-36]. When graphene is placed adjacent to $WSe_2$, proximity coupling generates two dominant SOC contributions: an Ising (valley-Zeeman) term that acts as an out-of-plane effective Zeeman field with opposite signs in the two valleys, and a Rashba term that favors in-plane spin polarization[33,35,37,38]. In this scenario, the Ising SOC opens a proximity-induced gap at each valley that carries finite Berry curvature and can therefore shift the valley Chern numbers, whereas Rashba SOC alone does not produce such topological modifications[39-41]. Crucially, symmetry considerations dictate that the relative weights of these two contributions depend on the crystallographic alignment angle Φ between $WSe_2$ and graphene[42-44]. For $WSe_2$, the Ising term is maximal near $\Phi \approx 0°$ and decreases almost linearly as Φ increases toward 30°, where it vanishes due to reflection symmetry that equalizes inter-valley and intra-valley tunneling amplitudes[42,44-46]. In contrast, the Rashba SOC is relatively insensitive to the twist angle and remains finite at $\Phi = 30°$[43-45]. This Φ-dependent competition between Ising and Rashba terms suggests that crystallographic alignment could serve as a tuning knob for the Berry curvature distribution and hence the topological character of moiré flat bands[36,46,47].

Proximity-induced SOC from TMDCs substantially modifies electronic properties in graphene heterostructures, influencing transport, symmetry-broken phases, and superconducting pairing[33,46,48-52]. In monolayer and bilayer graphene proximitized to $WSe_2$, weak antilocalization[37] and Landau level spectroscopy[38] have quantified both Ising and Rashba SOC strengths, and SOC-driven band inversion has been demonstrated in bilayer graphene[40]. In MATBG and TDBG interfaced with $WSe_2$, proximity SOC stabilizes superconductivity[48,50] and correlated phases[36], consistent with predictions that SOC modifies band topology[53]. Theory and recent experiments have established that the graphene–TMDC twist angle tunes the relative strengths of Ising and Rashba SOC in proximitized graphene[42-46]. For TDBG, whose flat bands acquire nontrivial valley Chern numbers under a displacement field[27,29,31,54], twist-angle control of SOC provides an additional handle for engineering correlated topological phases. However, how the competition between these SOC terms quantitatively determines the topological invariants of correlated insulating ground states, and how such control can be systematically achieved via twist-angle engineering, remains to be established.

Here we demonstrate that the crystallographic alignment angle Φ between $WSe_2$ and chiral-stacked twisted double bilayer graphene (cTDBG) deterministically controls whether the quarter-filled correlated ground state is a Chern insulator or topologically trivial. We focus on cTDBG for two reasons. First, unlike AB-AB stacked TDBG, the AB-BA configuration lacks $C_{2x}$ symmetry and therefore hosts flat bands with nonzero valley Chern numbers even at zero displacement field[27,29,55,56]. This nontrivial single-particle topology makes the system an ideal platform for investigating how proximity-induced SOC modifies Berry curvature distributions and correlated topological phases. Second, whereas the existing study of TDBG/$WSe_2$ heterostructures did not explicitly identify the stacking configuration[50], the interplay between proximity SOC and the intrinsic band topology of cTDBG remains unexplored. Using a continuum model that incorporates Φ-dependent Ising and Rashba SOC, we find that for $\Phi \approx 0°$ the Ising term dominates, inducing a spin-orbit gap with nonzero Berry curvature that renders the valence flat bands topologically nontrivial with valley Chern number $C_K = +2$. As Φ increases, the Ising contribution is suppressed by reflection symmetry and the bands evolve continuously toward a topologically trivial phase. Transport measurements on devices with well-characterized Φ confirm these theoretical predictions: at quarter filling, $\Phi \approx 0°$ devices exhibit a correlated Chern insulator with $C = +1$ and nearly quantized Hall resistance, whereas $\Phi \approx 15°$ devices display topologically trivial behavior with $C = 0$. These findings establish interlayer alignment as a practical and deterministic route to programming correlated topological phases in van der Waals moiré systems.

## Spin-orbit coupling effects on band topology in $WSe_2$/cTDBG

The system under study consists of cTDBG with monolayer $WSe_2$ interfaced to the top bilayer graphene, where carrier density $n$ and perpendicular displacement field $D$ serve as independent tuning parameters (Fig. 1**a**). The cTDBG comprises two Bernal-stacked bilayers with opposite chirality: the top bilayer has AB stacking while the bottom has BA stacking[56]. The system is characterized by two distinct twist angles: $\theta$ between the two graphene bilayers, and $\Phi$ between $WSe_2$ and the adjacent graphene layer. We employ $\theta \approx 0.9°$ to generate a moiré superlattice with triangular domains whose spatially varying interlayer registries shape the flat-band electronic structure (Fig. 1**b**, **c**). The alignment angle $\Phi$ governs the magnitude of proximity-induced SOC: the Ising SOC is maximized near $\Phi \approx 0°$ and vanishes at $\Phi = 30°$ due to symmetry, while the Rashba SOC remains relatively constant across all angles. This twist-angle dependence provides a tuning mechanism for the topological properties.

Proximity to monolayer $WSe_2$ induces substantial SOC in the adjacent graphene layer. Although first-principles calculations predict several symmetry-allowed SOC terms, including Ising, Rashba, Kane-Mele, and pseudospin-inversion-asymmetry (PIA) contributions, the low-energy band structure is governed primarily by the Ising and Rashba interactions[33,39,44,57]. The Kane-Mele term, while included in our model, is typically an order of magnitude weaker[33,58]. The PIA terms, though comparable in magnitude to Ising SOC, couple linearly to momentum and thus contribute negligibly to the flat-band physics[33,59,60]. Accordingly, the proximity-induced SOC comprises two primary contributions: an Ising term $H_I = \frac{\lambda_I}{2}\tau_z s_z$, where $\tau_z$ and $s_z$ are Pauli matrices acting on valley and spin degrees of freedom, respectively, and a Rashba term $H_R = \frac{\lambda_R}{2}(\tau_z \sigma_x s_y - \sigma_y s_x)$, where $\sigma$ acts on the sublattice. The Ising term acts as an effective out-of-plane Zeeman field with a valley-dependent sign, coupling the out-of-plane spin component to the valley index such that electrons in the $K$ valley are predominantly spin-up polarized while those in $K'$ are predominantly spin-down polarized[33,34,37]. This lifts the fourfold spin-valley degeneracy and splits each originally degenerate band into spin-polarized subbands, forming spin-valley-locked Kramers doublets. The Rashba term favours in-plane spin polarization and couples spin to the sublattice pseudospin, thereby inducing spin mixing. Both terms preserve time-reversal symmetry, which pairs states of opposite spin and valley, while producing opposite spin polarizations in the two valleys[39,58].

The relative magnitudes of $\lambda_I$ and $\lambda_R$ depend sensitively on the alignment angle $\Phi$, this dependence originates from virtual interlayer tunneling between graphene and $WSe_2$[33,35,38,39,42-45,57]. At $\Phi = 0°$, momentum conservation favours tunneling processes that couple each graphene valley predominantly to a single $WSe_2$ valley; the inherited spin-valley locking of $WSe_2$ then enhances the effective Ising SOC in graphene. At $\Phi \approx 30°$, the rotated Brillouin zones allow tunneling to the two $WSe_2$ valleys with comparable weight, leading to a substantial suppression of the net Ising contribution, such that Rashba SOC becomes the dominant mechanism. Our continuum model calculations, incorporating this orientation-dependent SOC, yield $|\lambda_I| \approx 2.0$ meV at $\Phi = 0°$ and a progressive reduction of $|\lambda_I|$ as $\Phi$ approaches 30°.

To elucidate how the $WSe_2$ orientation modulates band topology, we calculate the band structures along high-symmetry directions of the moiré Brillouin zone for representative alignment angles $\Phi = 0°$, 15°, and 30° at varying displacement fields $D$ (Fig. 1**d**–**l**). The line colour encodes the out-of-plane spin projection $\langle S_z \rangle$, and the dashed horizontal line denotes the chemical potential at charge neutrality point (CNP). At $\Phi = 0°$ (Fig. 1**d**–**f**), the bands exhibit pronounced spin splitting with well-defined $\langle S_z \rangle$ character over much of the moiré Brillouin zone: each originally spin-degenerate flat band is resolved into two largely spin-polarized subbands. This behaviour is consistent with dominant Ising SOC, which favours opposite out-of-plane spin orientations in the $K$ and $K'$ valleys. As $\Phi$ increases to 15° (Fig. 1**g**–**i**) and 30° (Fig. 1**j**–**l**), the spin splitting diminishes and $\langle S_z \rangle$ becomes less sharply defined, consistent with a growing Rashba contribution that mixes spin components.

The orientation-dependent SOC has pronounced consequences for band topology. Figure 2**j**–**m** presents $K$ valley band structures at fixed displacement field $D = 0.5$ V/nm, with Chern numbers labelled for the first valence and conduction bands for $\Phi = 0°$, 15°, 30°, and a reference case without $WSe_2$. For $\Phi = 0°$ (Fig. 2**j**), the first valence band carries $C_K$ = +2, whereas for Φ = 15° (Fig. 2**k**), Φ = 30° (Fig. 2**l**), and the reference structure without $WSe_2$ (Fig. 2**m**), we obtain $C_K$ = 0 for the same band at this displacement field. Time-reversal symmetry implies opposite Chern numbers in the $K'$ valley.

This topological transition can be understood as follows. In cTDBG, the reduced crystalline symmetry compared with AB-AB TDBG permits finite valley Chern numbers, although the actual topology depends on details of the band structure and displacement field. When Ising SOC dominates at Φ = 0°, it primarily splits bands into spin-resolved subbands without strongly perturbing the overall gap structure, preserving robust Berry curvature and enabling a nontrivial Chern assignment. By contrast, a stronger Rashba contribution introduces in-plane spin mixing that hybridizes states with different $\langle S_z \rangle$ character, which can close and reopen gaps near high-symmetry points and thereby redistribute Berry curvature, potentially changing the Chern numbers. The competition between these two SOC channels, tuned by Φ, thus controls whether the system resides in a topologically nontrivial or trivial regime.

The Fermi surface at quarter filling (ν = −1) provides additional insight into the SOC-induced modifications of Fermi surface geometry and band topology (Fig. 2**a**–**i**). We plot energy contours of the lower pair of flat valence bands below CNP with SOC included, and highlight the Fermi surfaces at ν = −1 (coloured lines) corresponding to the parameter combinations in Fig. 1**d**–**l**, with the colour indicating the in-plane spin projection. The out-of-plane spin projection along these surfaces is largely constant and can be inferred from Fig. 1**d**–**l**. Black dashed lines denote the corresponding Fermi surfaces without spin-orbit interaction. The large Fermi-surface deformation between the SOC and no-SOC cases, particularly evident at Φ = 0°, reflects the pronounced flatness of the bands near the Fermi energy: even meV-scale SOC splittings can qualitatively reshape the Fermi-surface geometry. This sensitivity to proximity effects is central to the orientation-tunable topological character demonstrated here.

## Transport signatures of correlated insulating states at quarter filling

The valley-contrasting Chern bands predicted for Φ ≈ 0° provide an ideal platform for exploring interaction-driven phases at partial band filling. To probe the ground states emerging from these flat bands, we performed transport measurements on a series of dual-gated $WSe_2$/cTDBG devices with $\Phi \approx 0°$ (see Methods for device fabrication). We focus on two representative devices with cTDBG twist angles of 0.93° and 0.80°, designated D1 and D2, respectively.

Fig. 3**a**, **b** displays the four-terminal longitudinal resistance $R_{xx}$ as a function of carrier density $n$ and displacement field $D$ for device D1 at $T = 1.6$ K and 20 mK in the absence of an applied magnetic field. The data reveal prominent resistance enhancements at quarter filling of the moiré bands. Both devices exhibit a robust correlated insulating state at $\nu = -1$, corresponding to quarter filling of the valence band, while device D2 additionally shows a correlated state at $\nu = +1$ in the conduction band. These resistive features strengthen markedly upon cooling, displaying insulating temperature dependence consistent with gapped correlated phases. The consistent observation of the $\nu = -1$ insulator across devices is particularly noteworthy: in previous TDBG studies without proximity-induced SOC, correlated states were observed exclusively in the conduction band owing to an intrinsic particle–hole asymmetry arising from the interlayer coupling. The emergence of robust valence-band correlations here suggests that SOC-induced band flattening and enhanced Berry curvature qualitatively modify the energetic landscape for correlated ground states.

The displacement-field dependence of the $\nu = -1$ feature is non-monotonic (Fig. 3**e**): the resistance initially increases with $D$ before decreasing at higher values. This behaviour reflects the combined influence of $D$ on the single-particle band structure and on interaction effects. Specifically, $D$ modifies the layer polarization of the low-energy wavefunctions and hence the effective proximity SOC experienced by carriers, tunes the flat-band dispersion and bandwidth, and can reshape the Berry curvature distribution in the moiré Brillouin zone. These effects together produce the observed non-monotonic evolution of the correlated insulating response.

To quantify the correlated gap, we measured the temperature dependence of $R_{xx}$ near $\nu = -1$ for device D2 at fixed $D = 0.6$ V/nm (Fig. 3f). The resistance at $\nu = -1$ strengthens progressively upon cooling and exhibits a characteristic minimum at $T^* \approx$ 2.7 K, marking a crossover from insulating behaviour ($dR/dT < 0$) at low temperatures to metallic behaviour ($dR/dT > 0$) at elevated temperatures. An Arrhenius fit to the activated regime below $T^*$ yields a thermal activation gap $\Delta = 0.78 \pm 0.04$ meV (Supplementary Fig. 6).

The emergence of a correlated insulating state at $\nu = -1$ is consistent with interaction-driven flavour symmetry breaking at quarter filling. In the fourfold spin-valley degenerate manifold, exchange interactions can favour a ground state in which carriers predominantly occupy a reduced set of flavours, thereby lowering the total energy through enhanced exchange. Such flavour polarization, if it selects a single valley, would break time-reversal symmetry and enable the occupied band to contribute its Chern number to transport. Whether the Chern character of the correlated state indeed tracks the single-particle band topology can be established through Landau-fan analysis and Hall measurements under applied magnetic field.

## Observation of orientation-tunable Chern insulators

To probe the topological character of the correlated states, we turn to magneto-transport measurements. Fig. 4**a**, **b** displays the longitudinal resistance $R_{xx}$ and Hall resistance $R_{xy}$ as a function of carrier density $n$ and perpendicular magnetic field $B$ for device D1 at $D = 0.6$ V/nm and $T = 25$ mK. The data reveal a Hofstadter-like Landau-fan structure with well-defined fans emanating from CNP, $\nu = -1$, and $\nu = -2$, as summarized schematically in Fig. 4**c**. From CNP, Landau fans with standard integer quantum Hall sequences emerge (black lines). Crucially, a Landau fan with slope corresponding to $C = +1$ (red lines) originates from the $\nu = -1$ correlated insulating state and becomes well developed above approximately 0.3 T, indicating that this state exhibits a nontrivial Chern response. Additional Landau fans with $C = +1$ and $C = +2$ (purple line) emanate from $\nu = -2$, consistent with sequential filling of Chern bands in the correlated regime.

The Hall resistance provides further characterization of the $\nu = -1$ state. Fig. 4**d** displays line cuts of $R_{xx}$ and $R_{xy}$ along the $C = +1$ trajectory emanating from $\nu = -1$. The Hall resistance approaches the quantized value $-h/e^2$, reaching 97.6% of quantization at $B \approx 3.0$ T, while the longitudinal resistance exhibits a pronounced minimum. This near-quantized Hall response, combined with the unit-slope Landau fan, establishes that the $\nu = -1$ correlated state exhibits a $C = +1$ Chern insulator response under applied magnetic field. The observed $C = +1$ is consistent with occupation of a single spin-split subband within one valley, in agreement with our theoretical prediction that each of the two predominantly spin-polarized subbands in the $K$ valley carries $C = +1$, summing to $C_K = +2$. This picture implies that the correlated ground state at $\nu = -1$ is flavour-polarized, occupying states predominantly from a single valley with nontrivial Chern character.

Our theoretical framework predicts that nontrivial band topology emerges only near $\Phi \approx 0°$, where Ising SOC dominates. To test this prediction, we examined device D3 with $\Phi \approx 15°$ and cTDBG twist angle $\theta = 0.83°$. The zero-field resistance map (Fig. 4**e**) was recorded at $T = 20$ mK, and the Landau fan diagram (Fig. 4**f**) at $T = 25$ mK, both well below 1.6 K, at which the $C = +1$ Landau fan is already clearly developed in $\Phi \approx 0°$ devices (Supplementary Fig. 4). The zero-field resistance map (Fig. 4**e**) exhibits correlated insulating signatures qualitatively similar to those observed in D1 and D2, with a clear resistance enhancement near $\nu = -1$. This observation demonstrates that the correlated insulating behaviour persists even in a regime where the single-particle bands are expected to be topologically trivial. Crucially, however, the Landau-fan diagram (Fig. 4**f**) reveals a striking contrast: the $\nu = -1$ feature shows no dispersion with magnetic field, corresponding to C = 0. This topologically trivial behaviour contrasts sharply with the $C = +1$ response observed at $\Phi \approx 0°$, consistent with the theoretical prediction that increasing Rashba SOC at larger $\Phi$ renders the bands topologically trivial.

The contrasting behaviour of $\Phi \approx 0°$ and $\Phi \approx 15°$ devices establishes the $WSe_2$ alignment angle as a control parameter for the topological character in this system. At $\Phi \approx 0°$, the dominant Ising SOC splits the flat bands into predominantly spin-polarized subbands with finite valley Chern numbers ($C_K = +2$, $C_{K'} = -2$). At quarter filling, the correlated insulating state exhibits a $C = +1$ Chern response under applied magnetic field, consistent with flavour polarization into a single valley with nontrivial Chern character. At $\Phi \approx 15°$, enhanced Rashba SOC renders the bands topologically trivial. Although the correlated insulating behaviour at $\nu = -1$ persists, as evidenced by similar zero-field transport signatures, the Landau-fan analysis yields $C = 0$, indicating that the topological response of the correlated state is governed by the single-particle band topology. These results demonstrate that correlated insulating behaviour at $\nu = -1$ occurs regardless of the band topology, while the Chern character of the resulting state is determined by the proximity-SOC-controlled topological properties of the underlying bands.

## Discussion and Outlook

Our key finding is that the crystallographic alignment angle $\Phi$ between $WSe_2$ and cTDBG governs whether the quarter-filled correlated ground state is topologically nontrivial or trivial. At $\Phi \approx 0°$, the $\nu = -1$ state exhibits a Chern insulator with $C = +1$ and near-quantized Hall resistance, whereas at $\Phi \approx 15°$ the same filling shows $C = 0$ despite comparable zero-field insulating behaviour. This contrast establishes crystallographic alignment as an independent tuning parameter for engineering correlated topological phases, complementary to twist angle, displacement field, and carrier density.

The underlying mechanism is qualitatively distinct from hBN-aligned moiré systems, where nontrivial topology arises from sublattice symmetry breaking[4,14,30]. In $WSe_2$/cTDBG, the topological character is instead governed by proximity-induced SOC whose relative strength depends on $\Phi$: Ising SOC dominates at $\Phi \approx 0°$ and endows each valley with finite Chern number, whereas emergent reflection symmetry at larger $\Phi$ suppresses the Ising contribution, driving the system into a topologically trivial regime. A systematic comparison of three scenarios (SOC only, sublattice asymmetry only, and both combined) confirms that the topological transition is driven predominantly by the $\Phi$-dependent Ising SOC, with sublattice symmetry breaking playing a subleading role (Supplementary Fig. 11; see Supplementary Section XI for details).

The robustness of the topological assignment is further supported by transport data from seven devices spanning three categories: three $\Phi \approx 0°$ devices (D1, D2, D4) all exhibiting $C = +1$, two large-$\Phi$ devices (D3 at $\Phi \approx 15°$, D5 at $\Phi \approx 30°$) both showing $C = 0$, and two bare cTDBG devices without $WSe_2$ (D6, D7) showing no correlated insulating state at $\nu = -1$ (Supplementary Table I, Supplementary Figs. 1–3). In particular, D2 and D5 share the same twist angle ($\theta = 0.80°$) and differ only in $\Phi$, providing a direct controlled comparison that isolates the crystallographic alignment as the determining variable. DFT calculations confirm that the dominance of Ising over Rashba SOC at $\Phi \approx 0°$ is preserved under realistic levels of strain, interlayer distance variation, and displacement field (Supplementary Figs. 9–10).

Crucially, correlated insulating behavior persists at both alignment angles, demonstrating that spontaneous flavor polarization at $\nu = -1$ does not require nontrivial single-particle topology; a nonzero Chern number arises only when the occupied flavor-polarized band carries net Berry curvature. Systematically mapping transport at intermediate alignment angles (for example, $\Phi \approx 5°$ and $10°$, near the predicted phase boundary $\Phi_c \approx 9°$) would clarify the nature and sharpness of the topological transition. The $|C| = 1$ character of the flavor-polarized bands motivates the search for fractional Chern insulators at fractional fillings, while achieving quantum anomalous Hall quantization at zero magnetic field[14,32,61], remains an open goal. More broadly, our results establish that crystallographic alignment constitutes a tuning knob that selects among topologically distinct correlated phases when Ising and Rashba SOC components compete, offering a route toward engineering the interplay between band topology and electron correlations.

## Methods

### Device fabrication

We fabricated the $WSe_2$/cTDBG devices using a twist-angle-controlled cut-and-stack dry transfer technique[56,62]. We first mechanically exfoliated monolayer $WSe_2$, bilayer graphene, few-layer graphite, and h-BN flakes onto $SiO_2$(300nm)/Si substrates, selecting high-quality flakes by optical and atomic force microscopy (AFM). The bilayer graphene was cut into two halves with an AFM tip prior to transfer. A dry-transfer stamp consisting of a poly(bisphenol A carbonate) (PC) film laminated onto a polydimethylsiloxane (PDMS) hemisphere on a glass slide was mounted on a micromanipulator. At 80 °C, the top graphite gate, top h-BN, and $WSe_2$ were sequentially picked up. We controlled the crystallographic alignment angle Φ between $WSe_2$ and bilayer graphene by matching the straight crystallographic edges of the two crystals under optical microscopy. To identify the zigzag crystallographic direction, we preferentially selected flakes presenting multiple long straight edges and used the inter-edge angles (60° or 120° between edges of the same family) to assign the crystallographic axes; the consistency of the measured Chern numbers across all aligned devices then provides an a posteriori confirmation of the alignment assignment. For Φ ≈ 0° devices, the edges were aligned as closely as possible; for Φ ≈ 15° and Φ ≈ 30° devices, a controlled in-plane rotation of the substrate stage was introduced after the initial alignment. The facet orientation measurements[63] indicate that exfoliated crystallographic edges exhibit an intrinsic angular spread of approximately 3° relative to the true crystallographic direction. Since our stacking procedure deliberately selects long, straight edges and aligns them under optical microscopy, the residual angular error in the assembled heterostructure is further reduced, and we conservatively estimate the alignment uncertainty in Φ to be approximately ±2°. This uncertainty is well below the theoretical topological phase boundary at $\Phi_c \approx 9°$ (Supplementary Fig. 11a), ensuring that the Φ ≈ 0° devices are unambiguously assigned to the topologically nontrivial regime, while the Φ ≈ 15° and Φ ≈ 30° devices lie within the trivial regime. Then, the first half of the bilayer graphene was then picked up at room temperature, after which the substrate stage was rotated by approximately 180.9° (i.e., 180° to account for the geometric flip plus the target twist angle of ~0.9°). The second half of the bilayer graphene was subsequently picked up at room temperature, completing the cTDBG stack. Finally, the bottom h-BN and bottom graphite gate were picked up at 80 °C, and the entire heterostructure was released onto a clean $SiO_2$/Si substrate at 140 °C.

One-dimensional edge contacts were fabricated by first defining the contact regions using electron-beam lithography, followed by reactive ion etching with a $CHF_3/O_2$ plasma (40/4 sccm, 4 Pa chamber pressure, 60 W radiofrequency power) to expose the graphene edges. Cr/Pd/Au (5/15/40 nm) was then deposited via standard electron-beam evaporation, with deposition rates of 0.1 Å $s^{-1}$ for Cr and Pd and 0.2 Å $s^{-1}$ for Au.

**Transport measurements**

Electrical transport measurements were performed in two cryogenic systems. Initial characterization at moderate temperatures was carried out in an Oxford cryostat with a base temperature of approximately 1.5 K. Resistance signals were measured using a Stanford Research SR830 lock-in amplifier operating at a frequency of 17.7 Hz. Gate voltages were applied using Keithley 2400 source meters. Low-temperature measurements were performed in a QONE dilution refrigerator with a base temperature below 20 mK, monitored by a factory-calibrated $RuO_x$ thermometer mounted on the mixing chamber plate. Top and bottom gate voltages were used to independently tune the carrier density $n$ and perpendicular displacement field $D$ according to $n = (C_{bg}V_{bg} + C_{tg}V_{tg})/e$ and $D = (C_{bg}V_{bg} - C_{tg}V_{tg})/2\varepsilon_0$, where $C_{tg}$ and $C_{bg}$ are the top and bottom gate capacitances per unit area, respectively, $e$ is the elementary charge, and $\varepsilon_0$ is the vacuum permittivity.

The twist angle $\theta$ of each device was determined from the carrier density $n_s$ corresponding to full filling of the moiré band (ν = ±4), using the relation $n_s = 8\theta^2/\sqrt{3}a^2$, where $a$ = 0.246 nm is the lattice constant of monolayer graphene.

**Continuum model of $WSe_2$/cTDBG**

We consider chiral-stacked twisted double bilayer graphene (cTDBG) consisting of an AB-stacked bilayer placed on top of a BA-stacked bilayer, twisted by an angle $\theta$ with respect to each other, with monolayer $WSe_2$ placed adjacent to the top graphene layer. The moiré superlattice constant is $L_s = a/2\sin(\theta/2)$, where $a$ = 0.246 nm is the lattice constant of monolayer graphene.

The low-energy electronic structure of cTDBG is described by the Bistritzer-MacDonald continuum model[5], in which the low-energy states from the two atomic valleys $K$ and $K'$ are decoupled. The continuum-model Hamiltonian for the $K$ valley is expressed as

$$H^K = \begin{pmatrix} H_{AB}^K & U^\dagger \\ U & H_{BA}^K \end{pmatrix} \tag{1}$$

where $H_{\mathrm{AB}}^K$ and $H_{\mathrm{BA}}^K$ are the Hamiltonians of the untwisted AB-stacked and BA-stacked bilayers, respectively:

$$H_{AB}^K = \begin{pmatrix} -\hbar v_F \boldsymbol{\sigma} \cdot (\boldsymbol{k} - \boldsymbol{K}_1) & h_+ \\ h_+^\dagger & -\hbar v_F \boldsymbol{\sigma} \cdot (\boldsymbol{k} - \boldsymbol{K}_1) \end{pmatrix} \tag{2}$$

$$H_{BA}^K = \begin{pmatrix} -\hbar v_F \boldsymbol{\sigma} \cdot (\boldsymbol{k} - \boldsymbol{K}_2) & h_- \\ h_-^\dagger & -\hbar v_F \boldsymbol{\sigma} \cdot (\boldsymbol{k} - \boldsymbol{K}_2) \end{pmatrix} \tag{3}$$

in which $\boldsymbol{K}_1(\boldsymbol{K}_2)$ denotes the Dirac point of the AB-stacked (BA-stacked) bilayer, and the Pauli matrices $\boldsymbol{\sigma} = (-\sigma_x, \sigma_y)$ act on the A, B sublattice degrees of freedom. The intralayer Fermi velocity is $v_F \approx 1.0 \times 10^6\ \mathrm{m\ s^{-1}}$.

The interlayer hopping matrix $h_+$ within the AB bilayer is

$$h_+ = \begin{pmatrix} -v_4 \pi^\dagger & \gamma_1 \\ -v_3 \pi & -v_4 \pi^\dagger \end{pmatrix} \tag{4}$$

where $\gamma_1 = 400$ meV is the direct interlayer coupling, and $v_3$, $v_4$ are the velocities associated with trigonal warping and electron-hole asymmetry[27,56], respectively, and $h_- = -h_+^\dagger$.

The interlayer coupling between the two twisted bilayers, connecting the top layer of the AB bilayer to the bottom layer of the BA bilayer, is expressed as

$$U = e^{-i\Delta \boldsymbol{K} \cdot \boldsymbol{r}} \mathcal{U}(\boldsymbol{r}) \tag{5}$$

where $\Delta \boldsymbol{K} = \boldsymbol{K}_2 - \boldsymbol{K}_1 = (0, 4\pi/3L_s)$ is the shift between the Dirac points of the two twisted bilayers. The moiré tunnelling term $\mathcal{U}(\mathbf{r})$ takes the form

$$\mathcal{U}(\boldsymbol{r}) = \begin{pmatrix} u_0 g(\boldsymbol{r}) & u_0' g(\boldsymbol{r} - \boldsymbol{r}_{AB}) \\ u_0' g(\boldsymbol{r} - \boldsymbol{r}_{AB}) & u_0 g(\boldsymbol{r}) \end{pmatrix} \tag{6}$$

where $\boldsymbol{r}_{\mathrm{AB}} = (\sqrt{3}L_s/3, 0)$, $g(\mathbf{r}) = \sum_{j=1}^{3} e^{i\boldsymbol{q}_j \cdot \boldsymbol{r}}$, and $u_0 \approx 79.7$ meV and $u_0' \approx 97.5$ meV are the intra-sublattice and inter-sublattice interlayer coupling terms, respectively. The dimer-site energy shift is set to $\Delta' = 50$ meV.

The displacement field $D$ is introduced by applying a linear electrostatic potential drop across the four graphene layers:

$$H_D^K = H_{AB\text{-}BA}^K + \text{diag}\left(-\frac{U_d}{2}, -\frac{U_d}{6}, \frac{U_d}{6}, \frac{U_d}{2}\right) \tag{7}$$

in which $U_d = eD \cdot d_t/\varepsilon$ is the electrostatic potential energy difference between the top and bottom layers, $d_t \approx 1.005$ nm is the total thickness of the four-layer system, and $\varepsilon \approx 3.0$ is the effective dielectric constant of the h-BN substrate.

**Proximity-induced spin–orbit coupling**

Proximity to monolayer $WSe_2$ induces spin–orbit coupling in the adjacent graphene layers. We incorporate four SOC contributions following refs[33,39,42,44].

The Ising (valley-Zeeman) term:

$$H_I = \frac{\lambda_I}{2}\tau_z s_z. \tag{8}$$

The Rashba term:

$$H_\text{R} = \frac{\lambda_\text{R}}{2}\left(\tau_z \sigma_x s_y - \sigma_y s_x\right). \tag{9}$$

The Kane-Mele (intrinsic) term:

$$H_\text{KM} = \frac{\lambda_\text{KM}}{2}\tau_z \sigma_z s_z. \tag{10}$$

The pseudospin-inversion asymmetry (PIA) term:

$$H_\text{PIA} = \frac{a}{2}[\lambda_{PIA}^A(\sigma_z + \sigma_0) + \lambda_{PIA}^B(\sigma_z - \sigma_0)]\left(k_x s_y - k_y s_x\right). \tag{11}$$

where $\tau_z$, $\sigma_i$, and $s_i$ are Pauli matrices acting on valley, sublattice, and spin degrees of freedom, respectively. For our calculations, we adopt representative proximity-SOC parameters $\lambda_\text{I}$ = 2.0 meV and $\lambda_\text{R}$ = 0.5 meV, based on first-principles calculations[35] for bilayer graphene/$WSe_2$, consistent with published monolayer DFT values[33,44,57], experimental measurements[37,38], and our own DFT calculations on graphene/$WSe_2$ heterostructures (Supplementary Figs. 9–10). We set the intrinsic (Kane–Mele) term to $\lambda_\text{KM}$ = 0.01 meV, and neglect the PIA contribution ($\lambda_\text{PIA} \approx 0$).

**Alignment-angle dependence of SOC**

The crystallographic alignment angle Φ between $WSe_2$ and graphene modulates the effective SOC parameters through momentum-conserving virtual tunnelling processes between the two materials. Following the symmetry analysis of ref.[44], the Ising and Rashba SOC exhibit distinct angular dependencies:

$$\lambda_{\mathrm{I}}(\Phi) = \lambda_{\mathrm{I},0}\cos(3\Phi) \quad (12)$$

$$\lambda_{\mathrm{R}}(\Phi) \approx \lambda_{\mathrm{R},0} \quad (13)$$

At $\Phi = 0°$, momentum conservation couples each graphene valley predominantly to a single $WSe_2$ valley, and the inherited spin-valley locking maximizes the Ising SOC ($|\lambda_{\mathrm{I}}| \approx 2.0$ meV). At $\Phi = 30°$, the rotated Brillouin zones allow tunnelling to both $WSe_2$ valleys with comparable amplitude, leading to cancellation of the Ising contribution ($\lambda_{\mathrm{I}} \to 0$), while the Rashba SOC remains finite. This angular dependence enables geometric control over the relative weight of Ising versus Rashba SOC, and hence over the band topology.

In our model, the SOC is applied to the top two graphene layers (L3 and L4) with an exponential decay: the SOC strength on L3 is scaled by a factor of 0.35 relative to L4, reflecting the spatial extent of the proximity effect.

**Chern number calculation**

The valley Chern number $C_K$ is computed by integrating the Berry curvature over the moiré Brillouin zone:

$$C_K = \frac{1}{2\pi}\int_{\mathrm{mBZ}} \Omega(\boldsymbol{k})\ d^2\boldsymbol{k} \quad (14)$$

where $\Omega(\mathbf{k}) = -2Im\sum_{n<n_F}\sum_{m>n_F}\frac{\langle n|\partial_{k_x}H|m\rangle\langle m|\partial_{k_y}H|n\rangle}{(E_m-E_n)^2}$ is the Berry curvature summed over occupied bands. Time-reversal symmetry implies $C_{K'} = -C_K$. A momentum-space cutoff of $6k_\theta$, where $k_\theta = 4\pi\theta/(3a)$ is the moiré wavevector, was used to ensure numerical convergence.

## Data Availability

All data generated in this study are provided in the Source Data file. Source data are provided with this paper.

## Code Availability

The source codes used for the continuum-model band-structure and valley-Chern-number calculations are provided with this paper in the Supplementary Software file. The first-principles calculations were performed using the open-source Quantum ESPRESSO package.

## References

1 Cao, Y. *et al.* Correlated insulator behaviour at half-filling in magic-angle graphene superlattices. *Nature* **556**, 80-84 (2018). https://doi.org/10.1038/nature26154

2 Cao, Y. *et al.* Unconventional superconductivity in magic-angle graphene superlattices. *Nature* **556**, 43-50 (2018). https://doi.org/10.1038/nature26160

3 Balents, L., Dean, C. R., Efetov, D. K. & Young, A. F. Superconductivity and strong correlations in moiré flat bands. *Nat. Phys.* **16**, 725-733 (2020). https://doi.org/10.1038/s41567-020-0906-9

4 Chen, G. *et al.* Tunable correlated Chern insulator and ferromagnetism in a moiré superlattice. *Nature* **579**, 56-61 (2020). https://doi.org/10.1038/s41586-020-2049-7

5 Bistritzer, R. & MacDonald, A. H. Moiré bands in twisted double-layer graphene. *Proc. Natl. Acad. Sci. U.S.A.* **108**, 12233-12237 (2011). https://doi.org/10.1073/pnas.1108174108

6 Andrei, E. Y. & MacDonald, A. H. Graphene bilayers with a twist. *Nat. Mater.* **19**, 1265-1275 (2020). https://doi.org/10.1038/s41563-020-00840-0

7 Lu, X. *et al.* Superconductors, orbital magnets and correlated states in magic-angle bilayer graphene. *Nature* **574**, 653-657 (2019). https://doi.org/10.1038/s41586-019-1695-0

8 Saito, Y., Ge, J., Watanabe, K., Taniguchi, T. & Young, A. F. Independent superconductors and correlated insulators in twisted bilayer graphene. *Nat. Phys.* **16**, 926-930 (2020). https://doi.org/10.1038/s41567-020-0928-3

9 Nuckolls, K. P. *et al.* Strongly correlated Chern insulators in magic-angle twisted bilayer graphene. *Nature* **588**, 610-615 (2020). https://doi.org/10.1038/s41586-020-3028-8

10 Pierce, A. T. *et al.* Unconventional sequence of correlated Chern insulators in magic-angle twisted bilayer graphene. *Nat. Phys.* **17**, 1210-1215 (2021). https://doi.org/10.1038/s41567-021-01347-4

11 Rozen, A. *et al.* Entropic evidence for a Pomeranchuk effect in magic-angle graphene. *Nature* **592**, 214-219 (2021). https://doi.org/10.1038/s41586-021-03319-3

12 Saito, Y. *et al.* Hofstadter subband ferromagnetism and symmetry-broken Chern insulators in twisted bilayer graphene. *Nat. Phys.* **17**, 478-481 (2021). https://doi.org/10.1038/s41567-020-01129-4

13 Saito, Y. *et al.* Isospin Pomeranchuk effect in twisted bilayer graphene. *Nature* **592**, 220-224 (2021). https://doi.org/10.1038/s41586-021-03409-2

14 Serlin, M. *et al.* Intrinsic quantized anomalous Hall effect in a moiré heterostructure. *Science* **367**, 900-903 (2020). https://doi.org/10.1126/science.aay5533

15 Wong, D. *et al.* Cascade of electronic transitions in magic-angle twisted bilayer graphene. *Nature* **582**, 198-202 (2020). https://doi.org/10.1038/s41586-020-2339-0

16 Xie, Y. *et al.* Fractional Chern insulators in magic-angle twisted bilayer graphene. *Nature* **600**, 439-443 (2021). https://doi.org/10.1038/s41586-021-04002-3

17 Zondiner, U. *et al.* Cascade of phase transitions and Dirac revivals in magic-angle graphene. *Nature* **582**, 203-208 (2020). https://doi.org/10.1038/s41586-020-2373-y

18 Yankowitz, M. *et al.* Tuning superconductivity in twisted bilayer graphene. *Science* **363**, 1059-1064 (2019). https://doi.org/10.1126/science.aav1910

19 Stepanov, P. *et al.* Untying the insulating and superconducting orders in magic-angle graphene. *Nature* **583**, 375-378 (2020). https://doi.org/10.1038/s41586-020-2459-6

20 Chen, G. *et al.* Signatures of tunable superconductivity in a trilayer graphene moiré superlattice.

Nature **572**, 215-219 (2019). https://doi.org/10.1038/s41586-019-1393-y

21 Xie, M. & MacDonald, A. H. Weak-Field Hall Resistivity and Spin-Valley Flavor Symmetry Breaking in Magic-Angle Twisted Bilayer Graphene. *Phys. Rev. Lett.* **127**, 196401 (2021). https://doi.org/10.1103/PhysRevLett.127.196401

22 Choi, Y. *et al.* Correlation-driven topological phases in magic-angle twisted bilayer graphene. *Nature* **589**, 536-541 (2021). https://doi.org/10.1038/s41586-020-03159-7

23 He, M. *et al.* Symmetry breaking in twisted double bilayer graphene. *Nat. Phys.* **17**, 26-30 (2021). https://doi.org/10.1038/s41567-020-1030-6

24 Cao, Y. *et al.* Nematicity and competing orders in superconducting magic-angle graphene. *Science* **372**, 264-271 (2021). https://doi.org/10.1126/science.abc2836

25 Han, T. *et al.* Signatures of chiral superconductivity in rhombohedral graphene. *Nature* **643**, 654-661 (2025). https://doi.org/10.1038/s41586-025-09169-7

26 Zhang, Y.-H., Mao, D., Cao, Y., Jarillo-Herrero, P. & Senthil, T. Nearly flat Chern bands in moiré superlattices. *Phys. Rev. B* **99**, 075127 (2019). https://doi.org/10.1103/PhysRevB.99.075127

27 Koshino, M. Band structure and topological properties of twisted double bilayer graphene. *Phys. Rev. B* **99**, 235406 (2019). https://doi.org/10.1103/PhysRevB.99.235406

28 Bultinck, N., Chatterjee, S. & Zaletel, M. P. Mechanism for Anomalous Hall Ferromagnetism in Twisted Bilayer Graphene. *Phys. Rev. Lett.* **124**, 166601 (2020). https://doi.org/10.1103/PhysRevLett.124.166601

29 Lee, J. Y. *et al.* Theory of correlated insulating behaviour and spin-triplet superconductivity in twisted double bilayer graphene. *Nat. Commun.* **10**, 5333 (2019). https://doi.org/10.1038/s41467-019-12981-1

30 Sharpe, A. L. *et al.* Emergent ferromagnetism near three-quarters filling in twisted bilayer graphene. *Science* **365**, 605-608 (2019). https://doi.org/10.1126/science.aaw3780

31 Shen, C. *et al.* Correlated states in twisted double bilayer graphene. *Nat. Phys.* **16**, 520-525 (2020). https://doi.org/10.1038/s41567-020-0825-9

32 Lu, Z. *et al.* Fractional quantum anomalous Hall effect in multilayer graphene. *Nature* **626**, 759-764 (2024). https://doi.org/10.1038/s41586-023-07010-7

33 Gmitra, M. & Fabian, J. Graphene on transition-metal dichalcogenides: A platform for proximity spin-orbit physics and optospintronics. *Phys. Rev. B* **92**, 155403 (2015). https://doi.org/10.1103/PhysRevB.92.155403

34 Wang, Z. *et al.* Origin and Magnitude of `Designer' Spin-Orbit Interaction in Graphene on Semiconducting Transition Metal Dichalcogenides. *Phys. Rev. X* **6**, 041020 (2016). https://doi.org/10.1103/PhysRevX.6.041020

35 Gmitra, M. & Fabian, J. Proximity Effects in Bilayer Graphene on Monolayer $WSe_2$: Field-Effect Spin Valley Locking, Spin-Orbit Valve, and Spin Transistor. *Phys. Rev. Lett.* **119**, 146401 (2017). https://doi.org/10.1103/PhysRevLett.119.146401

36 Lin, J.-X. *et al.* Spin-orbit–driven ferromagnetism at half moiré filling in magic-angle twisted bilayer graphene. *Science* **375**, 437-441 (2022). https://doi.org/10.1126/science.abh2889

37 Zihlmann, S. *et al.* Large spin relaxation anisotropy and valley-Zeeman spin-orbit coupling in $WSe_2$/graphene/h-BN heterostructures. *Phys. Rev. B* **97**, 075434 (2018). https://doi.org/10.1103/PhysRevB.97.075434

38 Wang, D. *et al.* Quantum Hall Effect Measurement of Spin–Orbit Coupling Strengths in Ultraclean Bilayer Graphene/$WSe_2$ Heterostructures. *Nano Lett.* **19**, 7028-7034 (2019).

https://doi.org/10.1021/acs.nanolett.9b02445
39 Gmitra, M., Kochan, D., Högl, P. & Fabian, J. Trivial and inverted Dirac bands and the emergence of quantum spin Hall states in graphene on transition-metal dichalcogenides. *Phys. Rev. B* **93**, 155104 (2016). https://doi.org/10.1103/PhysRevB.93.155104
40 Island, J. O. *et al.* Spin–orbit-driven band inversion in bilayer graphene by the van der Waals proximity effect. *Nature* **571**, 85-89 (2019). https://doi.org/10.1038/s41586-019-1304-2
41 Khatibi, Z. & Power, S. R. Proximity spin-orbit coupling in graphene on alloyed transition metal dichalcogenides. *Phys. Rev. B* **106**, 125417 (2022). https://doi.org/10.1103/PhysRevB.106.125417
42 Li, Y. & Koshino, M. Twist-angle dependence of the proximity spin-orbit coupling in graphene on transition-metal dichalcogenides. *Phys. Rev. B* **99**, 075438 (2019). https://doi.org/10.1103/PhysRevB.99.075438
43 David, A., Rakyta, P., Kormányos, A. & Burkard, G. Induced spin-orbit coupling in twisted graphene--transition metal dichalcogenide heterobilayers: Twistronics meets spintronics. *Phys. Rev. B* **100**, 085412 (2019). https://doi.org/10.1103/PhysRevB.100.085412
44 Naimer, T., Zollner, K., Gmitra, M. & Fabian, J. Twist-angle dependent proximity induced spin-orbit coupling in graphene/transition metal dichalcogenide heterostructures. *Phys. Rev. B* **104**, 195156 (2021). https://doi.org/10.1103/PhysRevB.104.195156
45 Sun, L. *et al.* Determining spin-orbit coupling in graphene by quasiparticle interference imaging. *Nat. Commun.* **14**, 3771 (2023). https://doi.org/10.1038/s41467-023-39453-x
46 Zhang, Y. *et al.* Twist-programmable superconductivity in spin–orbit-coupled bilayer graphene. *Nature* **641**, 625-631 (2025). https://doi.org/10.1038/s41586-025-08959-3
47 Bhowmik, S. *et al.* Spin-orbit coupling-enhanced valley ordering of malleable bands in twisted bilayer graphene on $WSe_2$. *Nat. Commun.* **14**, 4055 (2023). https://doi.org/10.1038/s41467-023-39855-x
48 Arora, H. S. *et al.* Superconductivity in metallic twisted bilayer graphene stabilized by $WSe_2$. *Nature* **583**, 379-384 (2020). https://doi.org/10.1038/s41586-020-2473-8
49 Zhang, Y. *et al.* Enhanced superconductivity in spin–orbit proximitized bilayer graphene. *Nature* **613**, 268-273 (2023). https://doi.org/10.1038/s41586-022-05446-x
50 Su, R., Kuiri, M., Watanabe, K., Taniguchi, T. & Folk, J. Superconductivity in twisted double bilayer graphene stabilized by $WSe_2$. *Nat. Mater.* **22**, 1332-1337 (2023). https://doi.org/10.1038/s41563-023-01653-7
51 Yang, J. *et al.* Impact of spin–orbit coupling on superconductivity in rhombohedral graphene. *Nat. Mater.* **24**, 1058-1065 (2025). https://doi.org/10.1038/s41563-025-02156-3
52 Patterson, C. L. *et al.* Superconductivity and spin canting in spin–orbit-coupled trilayer graphene. *Nature* **641**, 632-638 (2025). https://doi.org/10.1038/s41586-025-08863-w
53 Wang, T., Bultinck, N. & Zaletel, M. P. Flat-band topology of magic angle graphene on a transition metal dichalcogenide. *Phys. Rev. B* **102**, 235146 (2020). https://doi.org/10.1103/PhysRevB.102.235146
54 Burg, G. W. *et al.* Correlated Insulating States in Twisted Double Bilayer Graphene. *Phys. Rev. Lett.* **123**, 197702 (2019). https://doi.org/10.1103/PhysRevLett.123.197702
55 Kuiri, M. *et al.* Spontaneous time-reversal symmetry breaking in twisted double bilayer graphene. *Nat. Commun.* **13**, 6468 (2022). https://doi.org/10.1038/s41467-022-34192-x
56 Li, Q. *et al.* Tunable quantum criticalities in an isospin extended Hubbard model simulator.

Nature **609**, 479-484 (2022). https://doi.org/10.1038/s41586-022-05106-0

57 Zollner, K., João, S. M., Nikolić, B. K. & Fabian, J. Twist- and gate-tunable proximity spin-orbit coupling, spin relaxation anisotropy, and charge-to-spin conversion in heterostructures of graphene and transition metal dichalcogenides. *Phys. Rev. B* **108**, 235166 (2023). https://doi.org/10.1103/PhysRevB.108.235166

58 Kane, C. L. & Mele, E. J. Quantum Spin Hall Effect in Graphene. *Phys. Rev. Lett.* **95**, 226801 (2005). https://doi.org/10.1103/PhysRevLett.95.226801

59 Gmitra, M., Kochan, D. & Fabian, J. Spin-Orbit Coupling in Hydrogenated Graphene. *Phys. Rev. Lett.* **110**, 246602 (2013). https://doi.org/10.1103/PhysRevLett.110.246602

60 Bundesmann, J., Kochan, D., Tkatschenko, F., Fabian, J. & Richter, K. Theory of spin-orbit-induced spin relaxation in functionalized graphene. *Phys. Rev. B* **92**, 081403 (2015). https://doi.org/10.1103/PhysRevB.92.081403

61 Han, T. *et al.* Large quantum anomalous Hall effect in spin-orbit proximitized rhombohedral graphene. *Science* **384**, 647-651 (2024). https://doi.org/10.1126/science.adk9749

62 Chen, M. *et al.* Selective and quasi-continuous switching of ferroelectric Chern insulator devices for neuromorphic computing. *Nat. Nanotechnol.* **19**, 962-969 (2024). https://doi.org/10.1038/s41565-024-01698-y

63 Kolumbus, Y. et al. Crystallographic orientation errors in mechanical exfoliation. *J. Phys. Condens. Matter* **30**, 475704 (2018). https://doi.org/10.1088/1361-648X/aae877

## Acknowledgements

This work was supported in part by the National Key R&D Program of China under Grant 2025YFA1411003, 2023YFF0718400 and 2023YFF1203600, the National Natural Science Foundation of China (12322407, 62034004), the Natural Science Foundation of Jiangsu Province (No. BK20233001), the Leading-edge Technology Program of Jiangsu Natural Science Foundation (BK20232004), the Strategic Priority Research Program of the Chinese Academy of Sciences (XDB44000000), Innovation Program for Quantum Science and Technology (2024ZD0300101), Fundamental and Interdisciplinary Disciplines Breakthrough Plan of the Ministry of Education of China (JYB2025XDXM118, JYB2025XDXM120). F.M. and S.-J.L. would like to acknowledge support from the Fundamental Research Funds for the Central Universities (020414380227, 020414380240, 020414380242) and the e-Science Center of Collaborative Innovation Center of Advanced Microstructures. J.Y. and K.X. would like to acknowledge support from Zhejiang Provincial Natural Science Foundation of China under Grant No. LQN26A040023. K.W. and T.T. acknowledge support from the JSPS KAKENHI (Grant Nos. 21H05233 and 23H02052) and the World Premier International Research Center Initiative, MEXT, Japan.

## Author contributions

J.X., B.C. and F.M. conceived the idea and designed the experiments. F.M. and B.C. supervised the whole project. J.X. fabricated $WSe_2$/cTDBG devices and performed the measurements. J.X. and B.C. analyzed the experimental data. S.-J.L. participated in discussions of the project. Y.X., F.C., J.Y., S.C. and M.C. provided assistance in the experiments. J.Y. and K.X. set up the dilution refrigerator. J.X. developed the theoretical model. T.T. and K.W. provided h-BN samples. J.X., B.C. and F.M. wrote the manuscript with input from all authors.

## Competing interests

The authors declare no competing interests.

## Figures

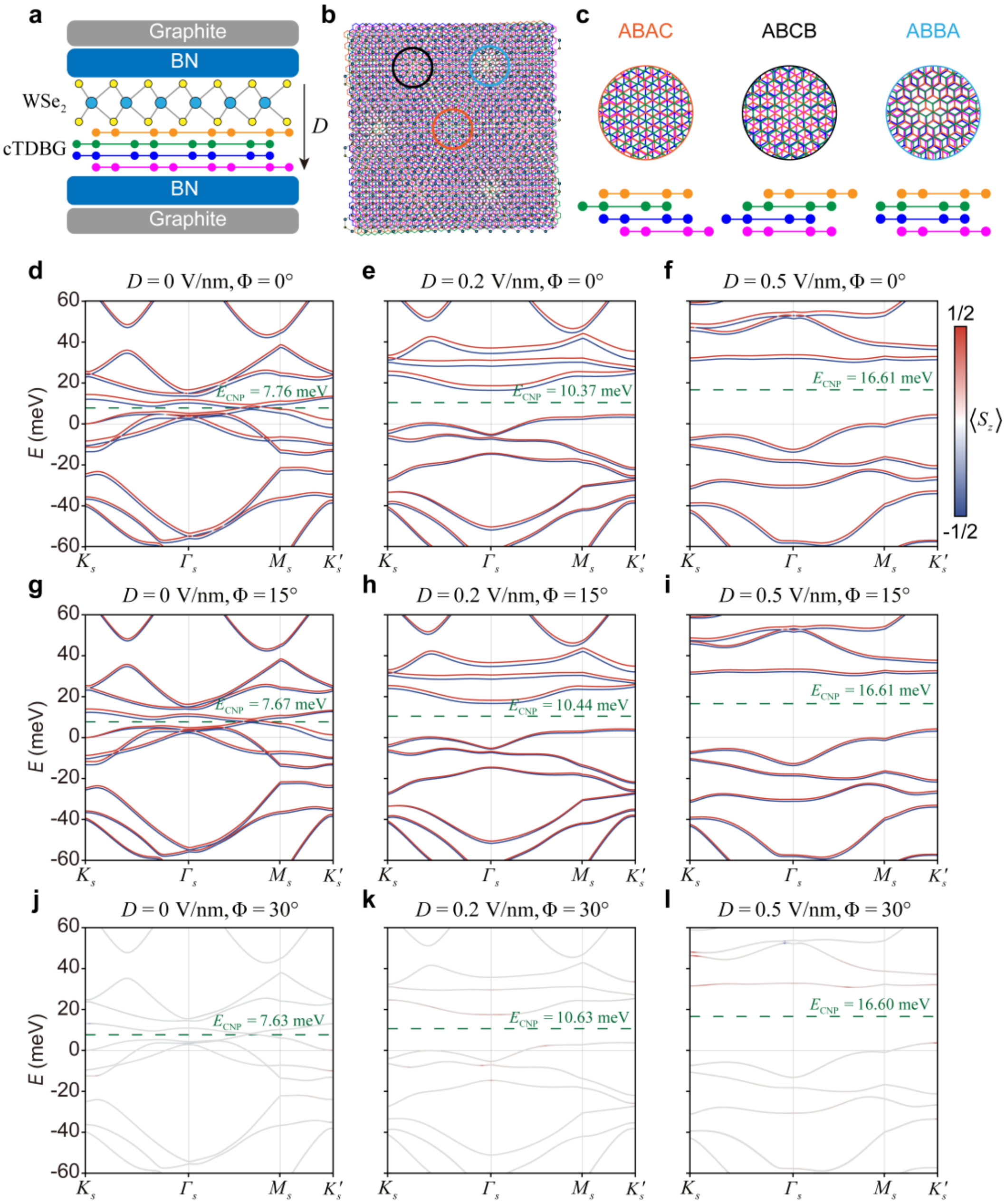

**Figure 1 | Device structure and band structure of $WSe_2$/cTDBG heterostructures.** **a**, Schematic illustration of the $WSe_2$/cTDBG stack, with hBN encapsulation (blue) and graphite top/bottom gates (grey) for independent control of carrier density $n$ and displacement field $D$. **b**, Moiré pattern of $WSe_2$-proximitized small-angle cTDBG, showing triangular domains with distinct stacking configurations (coloured circles). **c**, Top and side views of the three cTDBG stacking regions marked in **b** ($WSe_2$ layer omitted for clarity). **d–l**, Calculated band structures along high-symmetry directions of the moiré Brillouin zone for $WSe_2$ alignment angles $\Phi$ = 0° (**d–f**), 15° (**g–i**), and 30°

(**j–l**) at varying displacement fields $D$. The line colour represents the out-of-plane spin projection $\langle S_z \rangle$, and the dashed horizontal line denotes the chemical potential at CNP. The cTDBG twist angle is fixed at $\theta = 0.9°$.

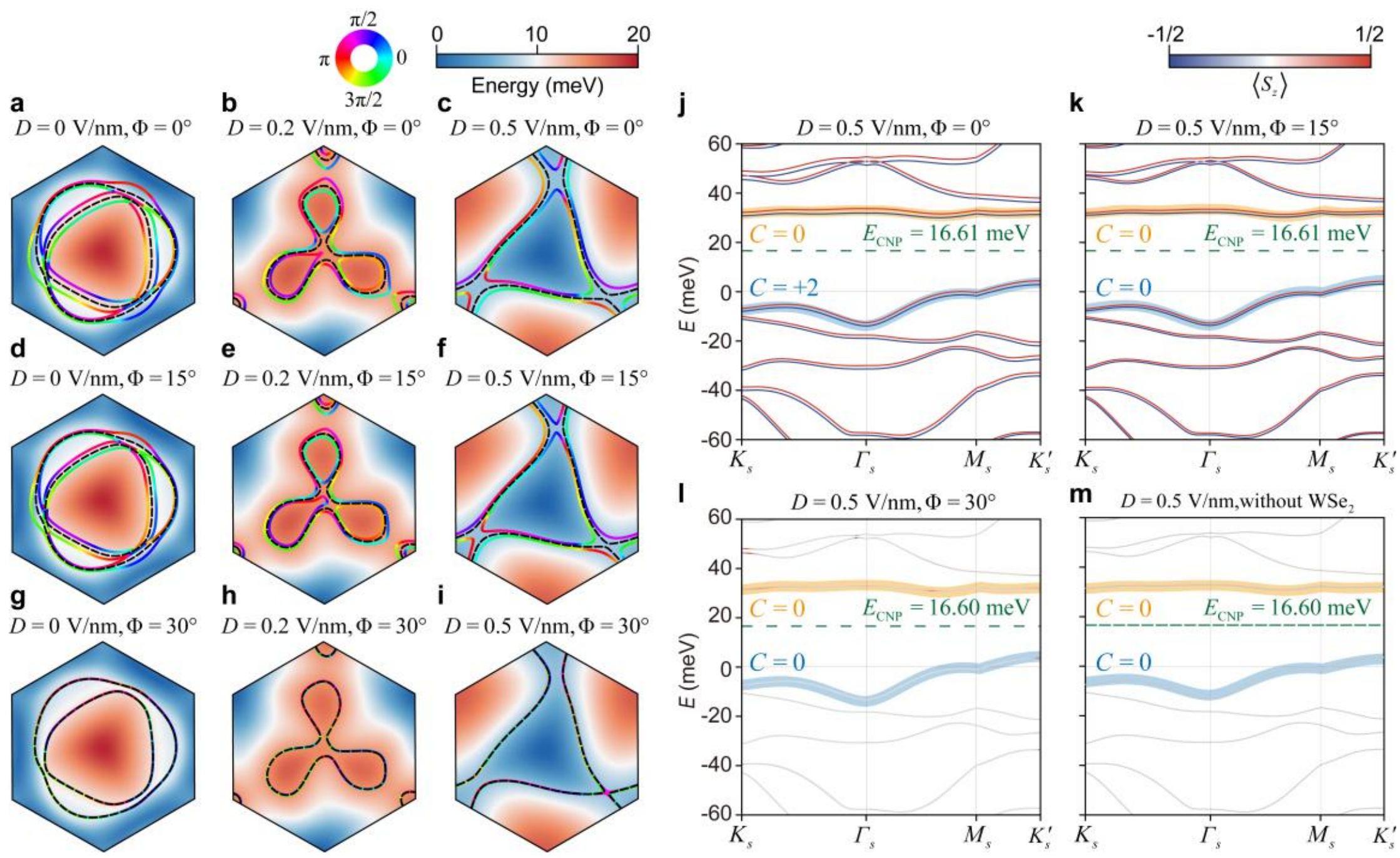

**Figure 2 | Fermi surfaces and valley Chern numbers. a–i**, Energy contours of the lower pair of flat valence bands below CNP with SOC included. Coloured lines show the Fermi surfaces at $\nu = -1$ corresponding to the parameter combinations in Fig. 1**d**–**l**, with the colour indicating the in-plane spin projection. The out-of-plane spin projection is largely constant along these surfaces and can be inferred from Fig. 1**d**–**l**. Black dashed lines denote the Fermi surfaces without spin-orbit interaction. The large Fermi-surface deformation reflects the pronounced flatness of the bands near the Fermi energy. **j**–**m**, K valley band structures at $D$ = 0.5 V/nm with Chern numbers labelled for the first valence and conduction bands: $\Phi = 0°$ (**j**, $C_K = +2$ for the first valence band, arising from two spin-split subbands each with $C = +1$); $\Phi = 15°$ (**k**, $C_K = 0$); $\Phi = 30°$ (**l**, $C_K = 0$); without $WSe_2$ (**m**, $C_K = 0$). Nontrivial topology emerges exclusively at $\Phi = 0°$ where Ising SOC dominates; the $K'$ valley carries opposite Chern numbers by time-reversal symmetry.

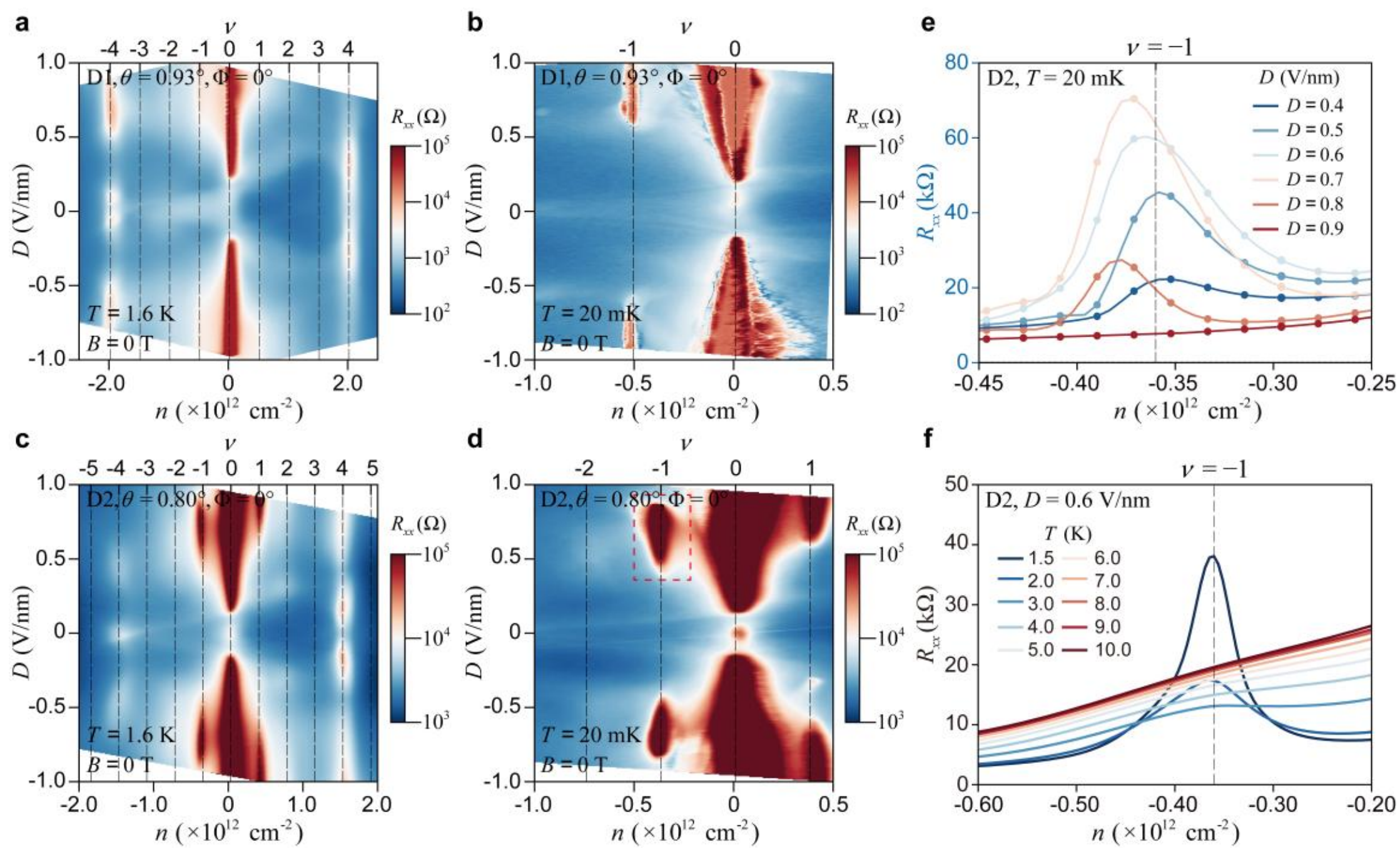

**Figure 3 | Zero-field transport revealing correlated insulating states. a**, **b**, Four-terminal longitudinal resistance $R_{xx}$ as a function of carrier density $n$ and displacement field $D$ for device D1 ($\theta = 0.93°$, $\Phi \approx 0°$) at $T = 1.6$ K (**a**) and 20 mK (**b**) in the absence of applied magnetic field. Top axis indicates moiré filling factor ν. **c**, **d**, Corresponding measurements for device D2 ($\theta = 0.80°$, $\Phi \approx 0°$). Dashed box highlights the correlated insulating state localized around ν = −1. **e**, Resistance line cuts near ν = −1 for D2, extracted from the dashed box region at varying displacement fields, showing non-monotonic $D$-dependence. The vertical dashed line indicates ν = −1 filling. **f**, Resistance line cuts near ν = −1 at varying temperatures for device D2 at $D = 0.6$ V/nm.

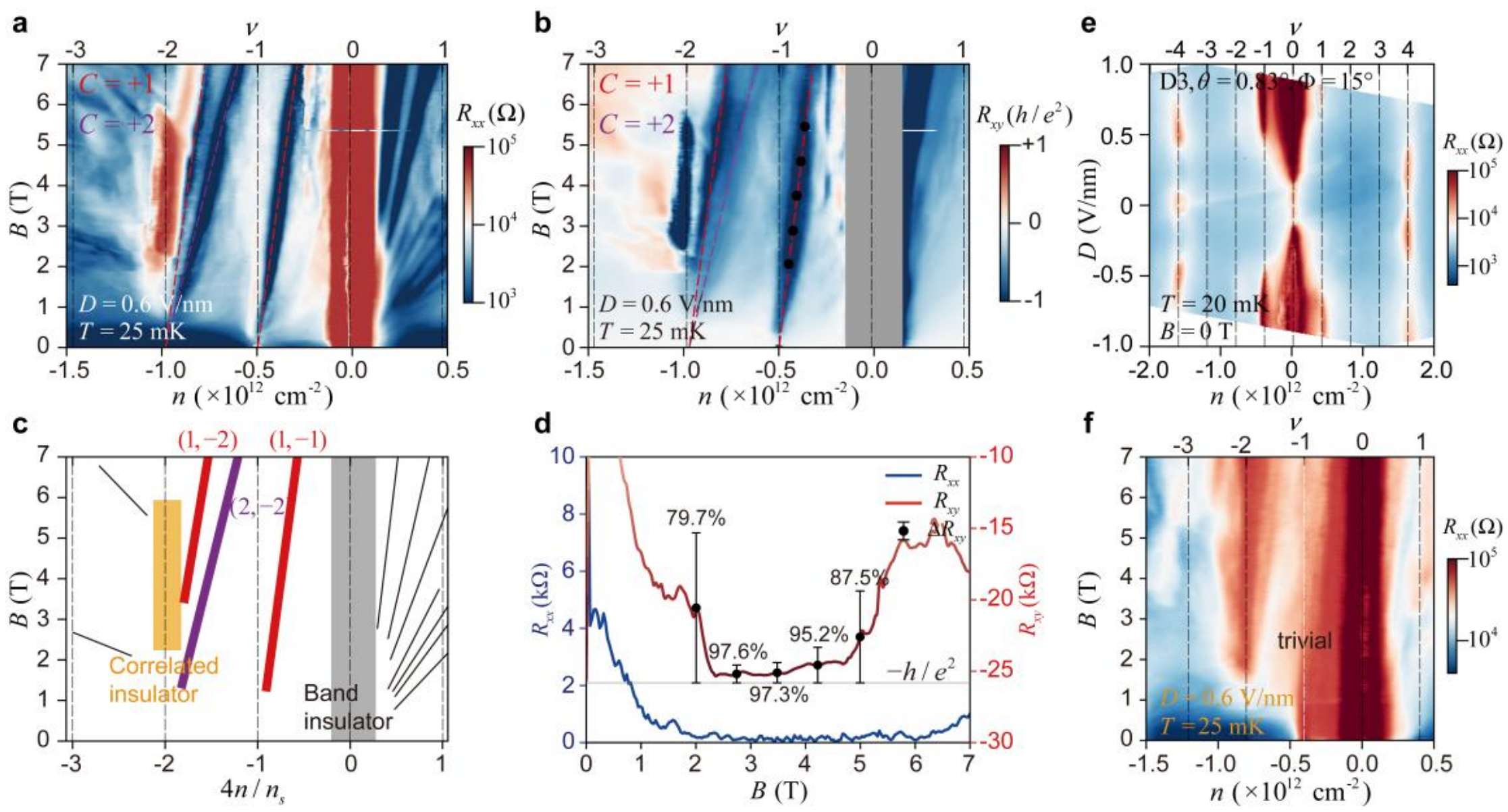

**Figure 4 | Landau fan diagrams and topological characterization.** **a**, **b**, Longitudinal resistance $R_{xx}$ (**a**) and Hall resistance $R_{xy}$ (**b**) for device D1 at $T = 25$ mK as a function of carrier density $n$ and perpendicular magnetic field $B$ at $D = 0.6$ V/nm, showing Hofstadter-like Landau-fan structure with fans emanating from CNP, ν = −1, and ν = −2. Red and purple dashed lines indicate slopes corresponding to $C = +1$ and $C = +2$, respectively. Grey shading near CNP in **b** masks regions where longitudinal admixture renders $R_{xy}$ unreliable. **c**, Schematic diagram of Landau fans in **a**. Coloured lines indicate Chern insulator states with (ν, $C$) = (−1, +1) and (−2, +1) (red), and (−2, +2) (purple); black lines emanate from CNP. **d**, Line cuts of $R_{xx}$ (blue, left axis) and $R_{xy}$ (red, right axis) along the $C = +1$ trajectory emanating from ν = −1, as indicated in **a** and **b**. Error bars on $R_{xy}$ indicate deviation from $-h/e^2$. Hall quantization reaches 97.6% of $-h/e^2$ at $B \approx 3.0$ T, consistent with a $C = +1$ Chern insulator response at ν = −1. **e**, Four-terminal longitudinal resistance $R_{xx}$ as a function of carrier density $n$ and displacement field $D$ for device D3 ($\theta = 0.83°$, $\Phi \approx 15°$) at $T = 20$ mK in the absence of applied magnetic field. Top axis indicates moiré filling factor ν. **f**, Longitudinal resistance $R_{xx}$ for device D3 at $T = 25$ mK as a function of carrier density $n$ and perpendicular magnetic field $B$ at $D = 0.6$ V/nm. The ν = −1 feature shows no dispersion with magnetic field ($C = 0$), confirming its topologically trivial character in contrast to $\Phi \approx 0°$ devices.